\documentclass[pre,twocolumn,superscriptaddress,amsmath,amssymb,noshowkeys,a4paper]{revtex4-1}

\usepackage[T1]{fontenc}
\usepackage{graphicx}
\usepackage{dcolumn}
\usepackage{bm}
\usepackage[usenames]{color}
\usepackage[dvipsnames]{xcolor}
\usepackage{epstopdf}
\usepackage[normalem]{ulem}
\usepackage[export]{adjustbox}

\newcommand{\affGRASP}{\affiliation{GRASP, Research unit CESAM, Institute of Physics B5a, Universit\'e de Li\`ege, Li\`ege, Belgium}}
\newcommand{\affErlangen}{\affiliation{PULS Group, Department of Physics, FAU Erlangen-N\"urnberg, IZNF, Erlangen, Germany}}
\newcommand{\affErlangenEng}{\affiliation{Deptartment of Chemical and Biological Engineering \& Department of Physics, FAU Erlangen-N\"urnberg, N\"urnberg, Germany}}
\newcommand{\affNurnberg}{\affiliation{Helmholtz Institute Erlangen-N\"urnberg for Renewable Energy (IEK-11), Forschungszentrum J\"ulich, N\"urnberg, Germany}}
\newcommand{\affZagreb}{\affiliation{Group for Computational Life Sciences, Ru{\dj}er Boskovic Institute, Zagreb, Croatia}}

\begin{document}
\title{The Scallop Theorem and Swimming at the Mesoscale}

\author{M. Hubert}
\thanks{These authors contributed equally}
\affErlangen
\author{O. Trosman}
\thanks{These authors contributed equally}
\affErlangen
\author{Y. Collard}
\thanks{These authors contributed equally}
\affGRASP
\author{A. Sukhov}
\thanks{These authors contributed equally}
\affNurnberg
\author{J. Harting}
\affNurnberg
\affErlangenEng
\author{N. Vandewalle}
\affGRASP
\author{A.-S. Smith}
\email{Correspondence to: maxime.hubert@fau.de, and \\ ana-suncana.smith@fau.de or asmith@irb.hr}
\affErlangen
\affZagreb

\preprint{\color{NavyBlue} DRAFT v11 - \today}


\begin{abstract}
By synergistically combining modeling, simulation and experiments, we show that there exists a regime of self-propulsion in which the inertia in the fluid dynamics can be separated from that of the swimmer. This is demonstrated by the motion of an asymmetric dumbbell that, despite deforming in a reciprocal fashion, self-propagates in a fluid due to a non-reciprocal Stokesian flow field. The latter arises from the difference in the coasting times of the two constitutive beads. This asymmetry acts as a second degree of freedom, recovering the scallop theorem at the mesoscopic scale.
\end{abstract}
\maketitle




The time-reversibility and linearity of the Stokes equation require microswimmers to deform in a non-reciprocal fashion in order to swim, a rule known as the \emph{scallop theorem} \cite{Purcell1977}. Many strategies in the Stokesian regime, requiring at least two degrees of freedom for successful propulsion, have been intensively investigated in the past decades \cite{Najafi2004,Ajdari2008,Avron2005,Ogrin2008,Pande2015,Ziegler2019,Dreyfus2005,Hamilton2017,Tierno2008}. This provided a fundamental understanding of the underlying dynamics as reflected by the emergence of several technological applications \cite{Leulmi2015,Gao2018,Medina-Sanchez2016,Xu2018,Orozco2015,Campuzano2012}.

\begin{figure}[!t]
	\includegraphics[width=0.9\columnwidth]{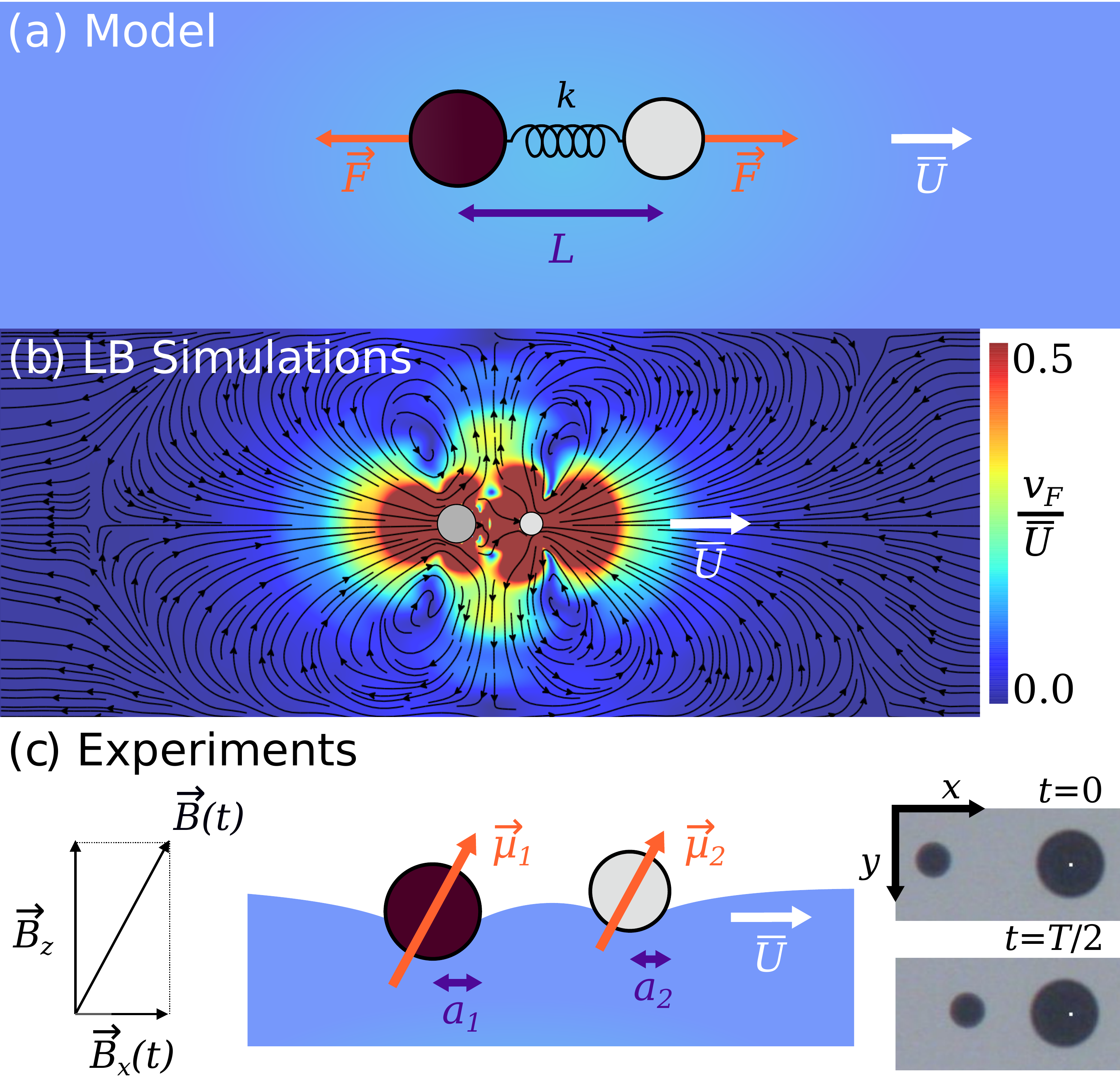}
	\caption{\textbf{Asymmetric swimming dumbbell.} 
	(a) The model assumes two spheres of density $\rho_s$ with radii $a_1$ and $a_2$ at a distance $L$ connected by a linear spring with constant $k$. The device is submerged in a fluid of viscosity $\eta$ and driven by sinusoidal forces of same amplitudes $F$ and frequencies $\omega$ acting in opposite directions. 
	(b) An equivalent system is addressed by lattice Boltzmann simulations ($a_1 = 5$, $a_2= 8$, $\eta = 1/6$, $k = 1/50$, $L = 28$, $\rho_f= 1$, and $\rho_s = 8$, all expressed in lattice units \emph{l.u.}). The simulation box is discretized by $400\times 160\times 160$ lattice nodes. The background shows the flow field averaged over one cycle of the external sinusoidal forcing ($F = 0.1$, $\omega= 1.57\times 10^{-3}$).
	(c) A magneto-capillary dumbbell is made of two ferromagnetic beads (magnetic moments $\vec \mu_i$, radii $a_i \in (397,500,793)\,\rm{\mu m}$, and density $7830\,\rm{kg/m^3}$), pinned at the water/air interface. The two beads separation $L$ is about $1400\rm{\mu m}$, and sets by the balance of capillary attraction and magnetic dipole repulsion. The device is driven by an external magnetic field $\vec B(t) = B_z \vec{e}_z + \left(B_0 + b \sin\left(\omega t\right)\right)\vec{e}_x$, which induces small oscillations of the beads. Snapshots are from the beginning of the cycle and half way through.}
	\label{fig_1}
\end{figure}

A natural way to break down the \emph{scallop theorem} is by introducing inertia. This is commonly achieved by the inertial dynamics of the fluid \citep{Lauga2007,Klotsa2019}, here characterized by the Reynolds number $\mathrm{Re_f} = \rho_f L \overline{U}/\eta $ ($\rho_f$ and $\eta$ being the fluid density and viscosity, $L$ the swimmer body length, and $\overline{U}$ the average swimming speed). For example, this can be achieved by using steady streaming \cite{Klotsa2015,Dombrowski2019,Dombrowski2020}, generation of vortices \cite{Hu2003}, or turbulent flows \cite{Gazzola2014}. The possibilities for the exploitation of the swimmer's own inertia, however, are still subject to debate \cite{Gonzalez-Rodriguez2009}. Interestingly, recent experiments and simulations have shown that mesoscopic structures, i.e. 100$\rm{\mu m}$ up to $1\rm{cm}$ in scale, display coasting effects \cite{Lagubeau2016,Sukhov2019,Grosjean2016}, while generating fluid flows with a time-reversible behavior \cite{Grosjean2015,Grosjean2018}. Those observations point to the possible existence of a swimming regime at low-$\mathrm{Re_f}$ where the inertia of the so-called mesoswimmer dominates and generates the motion, a hypothesis which warrants further investigations. 

A minimal mesoswimmer that can verify this hypothesis is an asymmetric dumbbell consisting of two different interacting beads, driven in a force free manner (Fig.~\ref{fig_1}). The Reynolds number $\mathrm{Re_s}$ of such a swimmer is set by its bead density $\rho_s$, its bead size $a$, and its beating frequency $\omega$ such as $\mathrm{Re_s} = \rho_s a^2 \omega/\eta$. This design possesses only one internal degree of freedom which leads to a reciprocal deformation, and therefore cannot swim without the help of inertia \cite{Klotsa2015,Dombrowski2019,Gonzalez-Rodriguez2009}. Assuming that the Reynolds number of the fluid $\mathrm{Re_f}\ll 1$ and of the swimmer $\mathrm{Re_s} \sim 1$, the flow should be dominated by the fluid viscosity while the propulsion mechanism should be related to the coasting time of the swimmer, which we define as $\tau = m/(6\pi\eta a)$, with $m$ being the bead mass. In this case, the swimmer should achieve propulsion and fulfill the requirement of the \emph{scallop theorem} by relying on an asymmetry in coasting times of the constitutive beads.

To elaborate on this idea,  we first build an analytic theory that relates the swimming velocity and coasting times. We successfully compare the model to experiments and lattice Boltzmann simulations in which no assumptions are imposed, thereby verifying the hypothesis that there is a swimming regime in which the inertial effects in the swimming dynamics can be separated and the swimmer coasting time harnessed for propulsion.

	
Our modeling efforts revolve around a dumbbell (Fig.~\ref{fig_1}a), that consists of two submerged beads of mass $m_i$ and radii $a_i$. The beads are linked by a linear spring with stiffness $k$ and natural length $L$, capturing, within the harmonic approximation $\vec{G}_{i,j} = -k(\vert\vec{x}_i-\vec{x}_j\vert-L)$, possible direct interactions between beads. The external forcing $\vec{F}_{i}$ is a sinusoidal force applied to each bead with the same intensity $F$ and frequency $\omega$ in opposite directions to satisfy the force-free condition. The swimming dynamics of this object is studied using the equations of motion 
\begin{equation}
    \frac{\partial \pmb{x}}{\partial t} =\hat{M}(\pmb{x})\left[\pmb{F}(t) + \pmb{G}(\pmb{x}) - \hat{m}\frac{\partial^2 \pmb{x}}{\partial t^2}\right],
    \label{MainDE_forces}
\end{equation} 
where bold symbols account for concatenated vectors, e.g. $\pmb{x} = (\vec{x}_1,\vec{x}_2)$. In this equation, we assume a low-$\rm{Re_f}$ dynamics by using the mobility matrix $\hat{M}(\pmb{x})$. This matrix models hydrodynamic interactions with the Stokes drag (diagonal elements) and the Oseen tensor (off diagonal elements). The inertia of the beads is explicitly taken into account by a force $-\hat{m}(\partial^2 \pmb{x}/\partial t^2)$. The matrix $\hat{m}$ has $m_1$ and $m_2$ in its diagonal elements. 

This equation is solved (see \cite{SI} (Sect. I.A)) using a perturbative scheme \cite{Ziegler2019}. Assuming $F/(ka_i) \ll 1$ and $a_i/L \ll 1$, one obtains the period-averaged swimming speed 
\begin{align}
	\overline{U}^F &=\frac{3F^2\omega_0^4}{2k^2\overline{\theta}^2}\frac{a_1^2 a_2^2}{\left(a_1+a_2\right)^3 L^2}\nonumber\\
		&\times \frac{\omega\left(\theta_2-\theta_1\right)}{\left(\left(\omega_0^2-\omega^2 + \frac{\omega^2}{\theta_1\theta_2}\right)^2 +\left(\frac{\omega_0^2}{\overline{\theta}}-\frac{\omega^2}{\theta_1}-\frac{\omega^2}{\theta_2}\right)^2\right)},
	\label{eq:Uf}
\end{align}
where $\theta_i = m_i\omega/(6\pi\eta a_i) = \tau_i\omega$ is the ratio of the coasting time to the external forcing period, $\overline{\theta} = (m_1+m_2)\omega/(6\pi\eta(a_1+a_2))$ is the swimmer coasting time, and $\omega_0^2 = k(m_1+m_2)/(m_1m_2)$. The solution for arbitrary separation (within the limit of the validity of the Oseen tensor) is provided in \cite{SI} (Sect.~I.A). The superscript $F$ in Eq.(\ref{eq:Uf}) refers to a force-based approach \cite{Pande2015,Pande2017,Ziegler2019} where the stroke of the beads is known only \emph{a posteriori}. 

Alternatively, one can impose a stroke \emph{a priori} and calculate the swimming velocity $\overline{U}^{S}$\cite{Najafi2004,Ajdari2008}. Now $\pmb{G}(\pmb{x})$ is removed from Eq.(\ref{MainDE_forces}). Assuming $a_i/L \ll 1$ and a stroke $\vert\vec{x}_2(t)-\vec{x}_1(t)\vert = L+d\sin(\omega t)$ (see \cite{SI} (Sect. I.B)), one obtains
\begin{equation}
	\overline{U}^{S} =  \frac{3d^2}{2}\frac{a_1^2 a_2^2}{\left(a_1+a_2\right)^3L^2} \,\frac{\omega\left(\theta_2-\theta_1\right)}{1+\overline{\theta}^2}.
	\label{eq:Us}
\end{equation}
Notably, there is a unique mapping between the two approaches (see \cite{SI} (Sect.~I.C)). 

In both force-based and stroke-based protocols, the analytical model described with Eq.~(\ref{MainDE_forces}) predicts a translation of the device in the direction of the beads with the smallest coasting time. This result may be sensitive to the $a_i/L$ conditions, as it can be seen in \cite{SI} where $\overline{U}^F$ is calculated without approximation beyond the use of the Oseen tensor. 

\begin{figure*}[!t]
	\centering
	\includegraphics[width=0.9\textwidth]{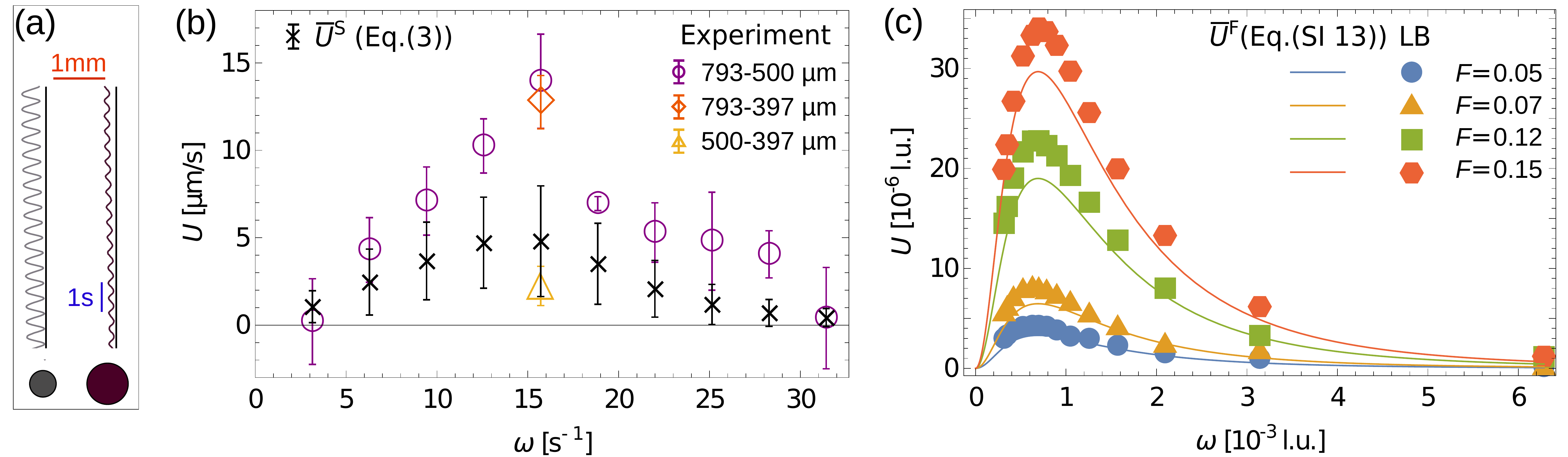}
	\caption{\textbf{Swimming dynamics of an asymmetric dumbbell - comparison of LB simulations and experiments with the analytic model} 
	(a) Sinusoidal trajectory of the beads ($500$ and $800{\rm \mu m}$) subject to the driving $B_z = 5.6 \,{\rm mT}$, $B_0= 0.7\,{\rm mT}$, $b = 0.35\,{\rm mT}$ and $\omega=12.57\,{\rm Hz}$. Vertical lines are guides to the eye. 
	(b) Average swimming speed of the asymmetric dumbbell as a function of the magnetic field frequency $\omega$. Amplitudes of the field are as in (a). Error bars account for the variance  ($\pm\sigma$) between 5 independent experiments. Theory using Eq.(\ref{eq:Us}) is shown in black for the $500-793{\rm \mu m}$ beads combination, with the error bars propagated from experimental uncertainty. 
	(c) Average swimming speed of the asymmetric dumbbell as a function of the frequency from lattice Boltzmann simulations (symbols) and the analytic model (lines) for different values of the driving force $F$ (no fitting). Parameters are same as in Fig.~\ref{fig_1} except that $k = 1/200$.  
	}
	\label{fig_2}
\end{figure*}


One can relax the assumptions made on the hydrodynamic flows and study the asymmetric dumbbell with lattice Boltzmann (LB) simulations (Fig.~1b) \cite{Benzi1992,Kruger2017,Sukhov2019}. This algorithm solves a discrete version of the Boltzmann equation and recovers solutions of the Navier-Stokes equations in the limit of low Mach and low Knudsen numbers. For the bead dynamics, a leap-frog algorithm is used to solve Newton's equation of motion. The beads are discretized on the fluid lattice and their dynamics is coupled to the fluid by a mid-grid bounce-back boundary condition \cite{Ladd2001,Harting2008,Sukhov2019,SI}. As such, both the fluid and the spring-connected beads are simulated without any dynamical assumptions (see \cite{SI} (Sect.~II.A)). For further comparison, we choose the numerical parameters to recover the expected Reynolds numbers of the beads and the fluid. We first confirm that there is no net flow responsible for the swimmer's displacement, and that a symmetric dumbbell does not swim. Finally, we show that a reciprocal deformation of an asymmetric pair results in a translational motion of the device in the direction of the small bead, as predicted by the theory. 


Finally, we perform experiments using magneto-capillary swimmers (see \cite{SI} (Sect.~II.B)) established previously \cite{Grosjean2016,Grosjean2018,Grosjean2015,Collard2020} (Fig.~\ref{fig_1}c). In short, the paramagnetic beads with a radius of 397, 500 or 793 $\rm{\mu m}$ are deposited on an air-water interface. When placed in a magnetic field $B_z$ perpendicular to the interface, their capillary attraction is balanced by magnetic dipole repulsion \cite{Grosjean2015}. Imposing a small oscillating field $\left(B_0 + b \sin\left(\omega t\right)\right)$ in the direction parallel to the interface induces oscillations in the relative distance between the beads. The homogeneity of $B_z$ and the flatness of the interface away from the beads ensure force-free conditions at all times. Consequently, symmetric dumbbells with two identical beads show no self-propulsion. However, a translation of the device is observed for two beads of different sizes. The swimmer moves towards the small bead, as shown in Fig.~\ref{fig_2}a (see also \cite{SI} SI Movie 1), in agreement with simulations and theoretical predictions. The swimmer is typically slow, reaching speeds up to $15\,\rm{\mu m/s}$, i.e. $4\times 10^{-3} L/T$ body-length $L$ per period $T$, which gives a flow dominated by viscous drag instead of inertia, as quantified with $\rm{Re}_f \sim 10^{-2}$. Similar speeds and Reynolds numbers were obtained in previous experiments involving the linear 3-bead swimmer \cite{Grosjean2016}.


In order to understand the role of the swimmer inertia, we analyze the frequency response of the swimming speed of the dumbbell (Fig.~\ref{fig_2}b,c) using all three approaches, whereby the parameters of the simulations are adjusted to recover the experimental swimmer geometry. In experiments, the investigated frequency range corresponds approximately to $\rm{Re_s} \sim 1.5$ up to $\rm{Re_s} \sim 15$, with the radius of the small bead used as the characteristic length (Fig.~\ref{fig_2}b). This is matched in simulations where $\rm{Re_s}$ ranges from $0.38$ to $7.54$ for the frequencies considered (Fig.~\ref{fig_2}c), while $\rm{Re_f}$ remains small at $\rm{Re_f} < 2\times 10^{-3}$. 

For low frequencies corresponding to $\rm{Re_s} \ll 1$, with $\rm{Re_f} \ll 1$, the asymmetric swimmer obeys the usually-encountered Stokesian \emph{scallop theorem} for microswimmers \cite{Purcell1977}. Consequently, the dumbbell swims inefficiently. A vanishing swimming speed is also observed at high frequencies in all approaches as the amplitude of oscillation decreases too. The intermediate frequencies are characterized by a broad  peak in the dumbbell speed. Following the analytic model, this maximum should be associated with the mechanical resonance of the dumbbell. Specifically,  Eq.~(\ref{eq:Uf}) possesses an optimal swimming frequency close to $\omega_0$, a signature of the influence of the swimmer inertia. This maximum is thus by nature different to the optimum frequency occurring for purely Stokesian dynamics \cite{Pande2015,Pande2017}. In experiments the maximum appears at a frequency of around $15\,\rm{s^{-1}}$, which corresponds to the characteristic mechanical resonance identified previously \cite{Lagubeau2016,Grosjean2016}. In simulations, it occurs around $\rm{Re_s} = 1.5$, which corresponds well to $\omega_0 \sim 1.22\times 10^{-3}\,\rm{l.u.}$. 

Finally, we compare the analytic model directly with the experiments (Fig.~\ref{fig_2}b) and simulations (Fig.~\ref{fig_2}c). Rather than using Eq.(\ref{eq:Uf}), we use SI-Eq.(13) (no restriction on $a_i/L$), due to the proximity of the beads in the simulations. With no fitting parameters, the agreement is excellent, with the error not exceeding 10\%. The strongest deviations are found around the peak velocity, where the non-zero fluid inertia may play a small role \citep{Dombrowski2019,Dombrowski2020,Klotsa2019}. Furthermore, from reading out the stroke amplitude obtained in experiments, the measured velocities can be compared to the model using Eq.~(\ref{eq:Us}). Once again, a very good agreement is obtained with some differences in speed amplitudes at higher frequencies. This deviation is attributed to the presence of the interface and the non-linearity of the magneto-capillary potential, which are not captured by the model. 

\begin{figure}[!t]
	\centering
	\includegraphics[width=\columnwidth]{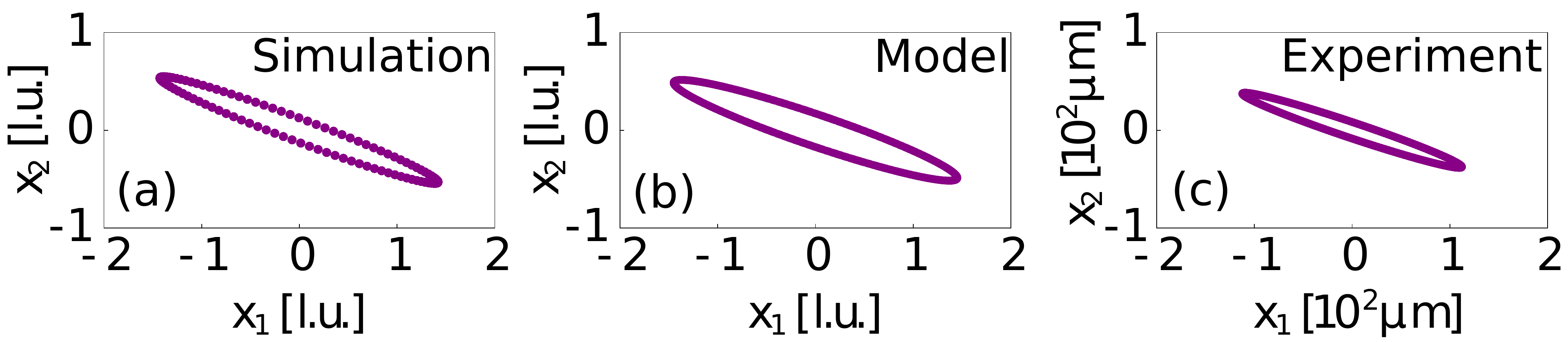}
	\caption{\textbf{Configuration space of the dumbbell.} The ellipse results from the phase shift in the oscillations of the beads $x_1(t)$ and $x_2(t)$ around the swimmer geometric center. (a)  LB simulations and (b) corresponding analytics  ($\omega = 1.57\times 10^{-3}$), and (c) experiments ($\omega = 15.7\,\rm{s^{-1}}$, 500-793$\rm{\mu m}$)}
	\label{fig_4}
\end{figure}

Those comparisons not only vindicate the theoretical model but also testify towards the existence of a mesoscopic swimming regime where propulsion is driven by  the inertia of the device while keeping a low $\rm{Re_f}$. Self-propulsion of mesoswimmers relies on $\rm{Re_s} > 1$, which points to the significant role of the inertia of the beads. However, as demonstrated by the behavior of the symmetric design, inertia alone is not able to propel with a reciprocal deformation. Indeed, swimming  necessitates the asymmetry of the design. Under the application of forces, beads accelerate and decelerate at a different rate as soon as $\theta_i \neq 0$ \cite{Gonzalez-Rodriguez2009}. A direct consequence of this asymmetric response is to induce a phase shift  (Fig.~\ref{fig_4}) in the oscillation of the beads measured within the laboratory frame (see \cite{SI} (Sect.~I.C)). This is captured by an ellipse in the configuration space of the dumbbell, spanned by the coordinates $\vec{x}_1$, and $\vec{x}_2$ of the two oscillating beads. As a consequence of this phase difference, the velocity of the beads with respect to the fluid is not  time reversible even though the swimmer deforms in a reciprocal fashion.  

The phase shift has two consequences. Firstly, it implies that despite having a force-free swimmer, the instantaneous flow-field generated by the swimmer can have a mono-polar component (see \cite{SI} Sect. I.D). Nevertheless, the time-averaged flow is dipolar (Fig.~\ref{fig_1}(b)), and the swimmer can be described as a puller in the investigated range of parameters. Secondly, knowing the phase-shift in the individual oscillations also allows us to cast the expression of the swimming speed into 
\begin{equation}
	\overline{U} \propto A_1 A_2 \sin\Delta\phi,
\end{equation}
where $A_i$ are the amplitudes of oscillation of the beads (see \cite{SI} (Sect.~II.B)). Both of these effects are consistent with the phase shift in the dynamics and flow fields generated in the simulations reported in \cite{Dombrowski2020}, where the dynamics of a dumbbell was investigated as a function of the fluid Reynolds number.

In conclusion, we used magneto-capillary swimmers and lattice Boltzmann simulations to provide the basis for a minimal theoretical model for swimming on the mesoscale. We show that there exists a dynamical regime where the swimmer inertia can be harnessed for self-propulsion in the low $\rm{Re_f}$ regime. Indeed, by including an asymmetry in coasting time in the design of the mesoswimmer, it is able to break the time-symmetry of the generated flow field.  The swimming velocity then is related to the area of the trajectory drawn in the configuration space. This  area is a measure of the non-reciprocity of the dynamics \citep{Ajdari2008,Purcell1977} and is typically used to demonstrate the \emph{scallop theorem}. The latter is, for the mesoswimmers, fulfilled by an intrinsic property of swimmer parts, namely their inertia that together mimic an independent degree of freedom. The analysis performed herein thus shows that the transition from  microswimmers to mesoswimmers may occur through a delicate balance of viscous damping and inertial relaxation. At higher Reynolds numbers, naturally, the inertia of the fluid will couple to the coasting of the swimmer and  dominate the dynamics. The analysis provided herein, however, may help to understand the emergence of this complex interplay.  

We thank S. Ziegler and G. Grosjean for insightful discussions. The funding was provided by DFG through the collaborative research center CRC1411 and the priority program SPP1726, as well by the FNRS grant PDR T.0129.18. Simulations were performed at the J\"ulich Supercomputing Centre, the High Performance Computing Center Stuttgart and the Regionales Rechenzentrum Erlangen.

\bibliographystyle{apsrev4-1}
\bibliography{2Beads}

\end{document}